\documentclass[prl, twocolumn, showpacs, amsmath, amssymb,superscriptaddress]{revtex4-1}
\usepackage{mysetup}
\newcommand{\syst}[2]{
\begin{tikzpicture}
    \def\efi{#1}
    \coordinate(A0) at (0.6cm,0cm);
    \foreach \a in {1,2,...,7}{
        \if\a1%
            \node[circle, minimum size=0.6cm, below right=\efi*\a and 1.2cm*\a of A0, thick](A\a){$\dots$};
        \else
            \if\a7%
                \node[circle, minimum size=0.6cm, below right=\efi*\a and 1.2cm*\a of A0, thick](A\a){$\dots$};
            \else
                \node[draw, circle, minimum size=0.6cm, below right= \efi*\a and 1.2cm*\a of A0, thick](A\a){};
            \fi
        \fi;
    }
    \foreach[evaluate=\a as \an using int(\a+1)] \a in {1,2,...,6}
        \path[]
            (A\a.north east) edge [
              text=black,
              thick,
              shorten <=2pt,
              shorten >=2pt,
              bend left=50
              ] node[above]{\IfEqCase{#2}{%
                              {1}{\(t\)}%
                              {0}{\(\)}%
                             }%
                           } (A\an.north west);
    \foreach[evaluate=\a as \an using int(\a+1)] \a in {1,2,...,6}
        \path[]
            (A\a.south east) edge [
              text=black,
              thick,
              shorten <=2pt,
              shorten >=2pt,
              bend right=50,
              dashed
              ] node[below]{\IfEqCase{#2}{%
                              {1}{\(U\)}%
                              {0}{\(\)}%
                             }%
                           } (A\an.south west);

    \foreach \a in {2,3,...,6}{
          \coordinate[below=2.6cm of A\a](r\a);
          \draw[thick, shading=axis, bottom color=white, top color=custBlue] (r\a)++(-0:0.4 and 1.5) arc(-0:180:0.4 and 1.5);
        \path[]
            (r\a)++(90:0.4 and 1.5) edge [
              text=black,
              thick,
              shorten <=2pt,
              shorten >=2pt,
              bend right=50
              ] node[left]{\IfEqCase{#2}{%
                              {1}{\(\Gamma\)}%
                              {0}{\(\)}%
                             }%
                           } (A\a.south);
    }
    \node[above=0.7cm of A2] (ar){};
    \draw[->] (ar) --node[above]{$E$} +(1.2cm,-\efi);
\end{tikzpicture}
}

\newcommand{\systDMRG}[2]{
\begin{tikzpicture}
    \def\efi{#1}
    \coordinate(A0) at (0.6cm,0cm);
    \foreach \a in {2,3,...,6}{
        \if\a1%
            \node[circle, minimum size=0.6cm, below right=\efi*\a and 1.2cm*\a of A0, thick](A\a){$\dots$};
        \else
            \if\a7%
                \node[circle, minimum size=0.6cm, below right=\efi*\a and 1.2cm*\a of A0, thick](A\a){$\dots$};
            \else
                \node[draw, circle, minimum size=0.6cm, below right= \efi*\a and 1.2cm*\a of A0, thick](A\a){};
            \fi
        \fi;
    }
    \foreach[evaluate=\a as \an using int(\a+1)] \a in {2,3,...,5}
        \path[]
            (A\a.north east) edge [
              text=black,
              thick,
              shorten <=2pt,
              shorten >=2pt,
              bend left=50
              ] node[above]{\IfEqCase{#2}{%
                              {1}{\(t\)}%
                              {0}{\(\)}%
                             }%
                           } (A\an.north west);
    \foreach[evaluate=\a as \an using int(\a+1)] \a in {2,3,...,5}
        \path[]
            (A\a.south east) edge [
              text=black,
              thick,
              shorten <=2pt,
              shorten >=2pt,
              bend right=50,
              dashed
              ] node[below]{\IfEqCase{#2}{%
                              {1}{\(U\)}%
                              {0}{\(\)}%
                             }%
                           } (A\an.south west);

    \foreach \a in {2,3,...,6}{
            \node[draw, circle, minimum size=0.6cm, below =1cm of A\a, thick](B\a){};
            \path[] (B\a) edge[
              text=black,
              thick,
              shorten <=2pt,
              shorten >=2pt,
              bend right=50
              ] (A\a);

    }

    \foreach[evaluate=\a as \an using int(\a+1)] \a in {2,3,...,5}
        \path[]
            (B\a.north east) edge [
              text=black,
              thick,
              shorten <=2pt,
              shorten >=2pt,
              bend left=50
              ] node[above]{\IfEqCase{#2}{%
                              {1}{\(t\)}%
                              {0}{\(\)}%
                             }%
                           } (B\an.north west);
\end{tikzpicture}
}

\begin{document}
\title{Non-equilibrium Properties of Berezinskii-Kosterlitz-Thouless Phase Transitions}

\author{C. Kl\"ockner}
\affiliation{Technische Universit\"at Braunschweig, Institut f\"ur Mathematische Physik, Mendelssohnstraße 3, 38106 Braunschweig, Germany}

\author{C.\ Karrasch}
\affiliation{Technische Universit\"at Braunschweig, Institut f\"ur Mathematische Physik, Mendelssohnstraße 3, 38106 Braunschweig, Germany}

\author{D.M.\ Kennes}
\affiliation{Institut f\"ur Theorie der Statistischen Physik, RWTH Aachen University and JARA-Fundamentals of Future Information Technology, 52056 Aachen, Germany}
\affiliation{Max Planck Institute for the Structure and Dynamics of Matter, Center for Free Electron Laser Science, 22761 Hamburg, Germany}
\begin{abstract} 
We employ a novel, unbiased renormalization-group approach to investigate non-equilibrium phase transitions in infinite lattice models. This allows us to address the delicate interplay of fluctuations and ordering tendencies in low dimensions out of equilibrium. We study a prototypical model for the metal to insulator transition of spinless interacting fermions coupled to electronic baths and driven out of equilibrium by a longitudinal static electric field. The closed system features a Berezinskii-Kosterlitz-Thouless transition between a metallic and a charge-ordered phase in the equilibrium limit. We compute the non-equilibrium phase diagram and illustrate a highly non-monotonic dependence of the phase boundary on the strength of the electric field: For small fields, the induced currents destroy the charge order, while at higher electric fields it reemerges due to many-body Wannier-Stark localization physics. Finally, we show that the current in such an interacting non-equilibrium system can counter-intuitively flow opposite to the direction of the electric field. This non-equilibrium steady-state is reminiscent of an equilibrium distribution function with an effective negative temperature.     
\end{abstract}

\pacs{} 
\date{\today} 
\maketitle

{\it Introduction---} Understanding the properties of quantum many-body systems which are {\it continuously} driven out of equilibrium is at the vanguard of contemporary condensed matter research \cite{Basov2017}. A particularly intriguing avenue along these lines is the study of the interplay of non-equilibrium physics with emergent phenomena such as phase transitions, which are well understood in equilibrium \cite{Sachdev2009}. Driving a system out of its equilibrium state in a controlled fashion in order to explore this interplay poses an experimental challenge, which is being mastered at an astonishing rate \cite{Budden20}. Reliable theoretical tools to connect to these experimental advances from the viewpoint of microscopic models are being complicated by the inherent complexity of treating a system in non-equilibrium. In low dimensions, thermal and quantum fluctuations are enhanced, which leads to a destabilization of many ordered phases of matter, especially those that break a continuous symmetry \cite{Mermin66,Hohenberg67,PhysRevB.100.121110}. Even in equilibrium, capturing this competition of fluctuations and order is well beyond simple mean-field approaches which inherently underestimate the relevance of fluctuations and which thus artificially promote long-ranged order. This is particularly prominent in one dimension where quantum and thermal fluctuations do not allow for spontaneously-broken continuous symmetries. There is a dire need to develop tools that aid an understanding of the non-equilibrium physics of low-dimensional systems beyond simple mean-field paradigms \cite{Jakobs2007,Mitra2008,Kennes2012,Aoki2014,Sieberer2016,AK1,PhysRevLett.120.127601}.

One of the most successful approaches to describe emergent phenomena in equilibrium is the so-called renormalization-group (RG), which has been instrumental in understanding and describing phase transitions of classical as well as quantum many-body systems \cite{RevModPhys.46.597,RevModPhys.47.773}. The key idea of the RG is to address energy scales successively (usually from high to low) in order to set up an effective low-energy theory that is easier to handle. In non-equilibrium a clear separation into high and low energies is ambiguous due to the continuous drive and as a consequence such an ansatz needs to correctly account for the {\it microscopic} details of the underlying model, which are often disregarded in an  equilibrium RG procedure. In this work, we employ an out-of-equilibrium RG scheme \cite{Metzner2012,kopietzBook} which does keep the full microscopic information about high- and low-energy degrees of freedom. In contrast to prior applications, we do account for inelastic two-particle scattering terms. Our approach is thus an \textit{a priori} well-suited candidate to explore the non-equilibrium realm.

We stress that in equilibrium and at low energies, the microscopic model studied in this paper can be mapped to a continuous field theory (the so-called Sine-Gordon model). Following this route is tremendously insightful, but requires care in an out-of equilibrium situation. One can directly study the Sine-Gordon model in non-equilibrium, but since the separation into high- and low-energy degrees of freedom is ambiguous in the presence of a continuous drive, the underlying mapping (from a microscopic model to a continuum field theory) itself becomes less clear. For a quenched system (a system that is not continuously driven but for which the Hamiltonian is abruptly changed in time), recent pioneering works on 2D superfluids  \cite{MatheyL2010} and on 2D cold atoms \cite{MatheyL2017} reveal Berezinskii-Kosterlitz-Thouless (BKT)-like behavior \cite{Kosterlitz_1973,giamarchi} in the time evolution {\it towards a steady-state}. Our study is complementary in spirit, as we i) directly work with the microscopic model and ii) analyze the {\it steady-state itself} which emerges for continuous driving. This state is an inherently non-equilibrium one; e.g., it violates detailed balance, as we will demonstrate later. Our results can serve as a test bed to determine if and how microscopic models can be mapped to field theories in a continuously-driven out-of-equilibrium setup.

On general grounds, a study of the non-equilibrium emergent behavior of quantum many-body systems needs to include the coupling to a bath. In a closed quantum system with genuine interactions and drive, heating is expected to push the system into a hot (asymptotically infinite-temperature) state where all interesting emergent phenomena are wiped out entirely. Therefore, continuously driving a system out of equilibrium necessitates another tuning knob -- the strength of reservoir-system coupling. This will fundamentally affect the physics. Our RG procedure can account for both the driving and the coupling to reservoirs.

Our strategy is to start with a system in which a well-understood BKT phase transitions occurs in equilibrium \cite{Kosterlitz_1973,giamarchi} and to then apply our novel framework to investigate the intriguing realm of non-equilibrium. We focus on a model of interacting lattice fermions which features a BKT transition from a metallic into a charge-ordered (insulating) state in equilibrium. Our RG calculations yield good agreement with this exact result as well as with mean-field predictions for the open system in limits where quantum fluctuations are irrelevant (e.g., for large interactions). This illustrates that our scheme is well-suited to determine the out-of-equilibrium properties in arbitrary parameter regimes. As a key result, we find that a longitudinal static electric field has a profound influence on the steady-state non-equilibrium phase diagram: A small field induces currents and can thus eventually suppress charge order and drive the system into a metallic state. For large fields, however, the current is blocked by many-body Wannier-Stark localization \cite{Wannier1962,Stark1,Stark2}: As particles move either up or down the lattice, they lose/gain energy from the linear potential gradient induced by the electric field and are finally reflected. The system consequently re-enters an insulating, charge-ordered state.  A second confounding observation is that the non-equilibrium steady-state can evolve into one of effective negative temperature \cite{Ramsey56,Mosk05,Braun52}. In such a state, the current flows opposite to the applied electric field \cite{PhysRevLett.91.056803,PhysRevB.71.115316}, which in equilibrium would signal the onset of an instability to phase separation of charge.

\begin{figure}
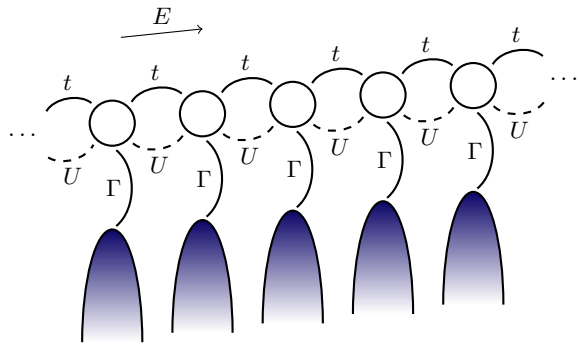

    \syst{-0.125}{1}
    \caption{ The model used in this paper: a tight-binding chain with a nearest-neighbor hopping \(t\) and a nearest-neighbor interaction \(U\) is coupled to electronic reservoirs via a hybridization \(\Gamma\). A longitudinal electric field $E$ drives the system out of thermal equilibrium. }\label{fig:pic_chain}
\end{figure}

{\it Model and Method--- }
\label{sec:model}
We consider an infinite interacting, one-dimensional metal in an static electric field:
\begin{equation}
\label{eq:modelHam}
\begin{split}
H=&\sum_{n} t c^\dagger_{n}c_{n+1} + {\rm h.c.}+nEc^\dagger_nc_n\\&+ U\left(c^\dagger_{n}c_{n}-\frac{1}{2}\right)\left(c^\dagger_{n+1}c_{n+1}-\frac{1}{2}\right),
\end{split}
\end{equation}
where $c^{(\dagger)}_{n}$ denote fermionic annihilation and creation operators. The strength of the nearest-neighbor hoppings and two-body interactions is given by $t$ and $U$, respectively. The electric field $E$ leads to an energy bias between adjacent sites. We side-couple every site $n$ of this chain to an individual electronic zero-temperature ($T=0$, chemical potential $\mu_n=nE$) reservoir whose bandwidth is assumed to be large compared to all other energy scales. The reservoirs are then fully characterized by a scalar hybridization \(\Gamma\). Our model is visualized in Fig.~\ref{fig:pic_chain}. 

In the absence of the reservoirs and in the equilibrium limit ($\Gamma=E=0$), the model can be solved using the Bethe ansatz \cite{giamarchi}. The ground state is a gapless Luttinger liquid for small $U$, but charge density wave order is stabilized beyond a {\it finite} critical $U_\mathrm{crit}/t=2$. The phase transition is driven by a Berezinskii-Kosterlitz-Thouless mechanism. Our goal is to determine the out-of-equilibrium phase diagram for arbitrary $\Gamma$ and $E$.

\begin{figure}
\includegraphics[width=\columnwidth,clip]{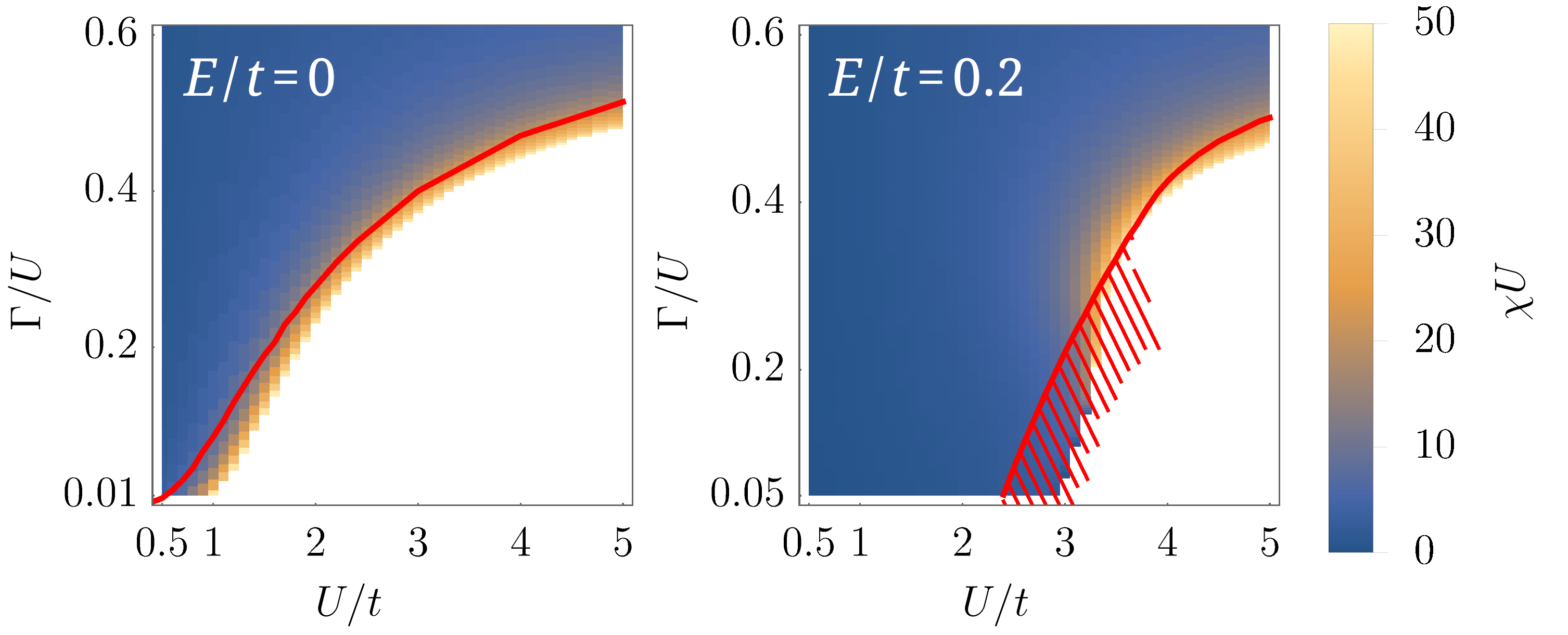}
  \caption{Phase diagram in the $U$-$\Gamma$-plane for two different electric fields $E$. The color scale displays the susceptibility \(\chi\) towards a charge density wave (white regions) obtained using our RG scheme \cite{longpaper}. It is known that the system is driven into a CDW phase at $U_\text{crit}/t=2$ in the absence of reservoirs in equilibrium ($\Gamma=E=0$). Our RG approach yields $U_\text{crit}^\text{RG}/t\approx1.4$ in this limit in contrast to mean-field theory (thick red line), which finds $U_\text{crit}^\text{MF}=0$. When a small electric field $E/t=0.2$ is switched on, currents start to flow and suppress charge order; thus, the size of the metallic phase increases. In certain areas of the phase diagram (red hatching), one finds multiple stable solutions of the mean-field equations (where at least one indicates charge order and at least one indicates a uniform state). 
  }\label{fig:inter_field}
\end{figure}

A mean-field analysis \cite{Kamenevbook,Mitra2008,PhysRevB.100.121110} for $\Gamma=E=0$ yields $U^{\rm MF}_\mathrm{crit}/t=0$ in stark contrast to the exact result. Critical quantum fluctuations are disregarded, and a mean-field approach is inherently insufficient to study the phase diagram. The same holds true for second order perturbation theory in $U$, which does not find a transition into a CDW phase at any $U$. Thus, a more sophisticated method is required. In this paper, we employ a novel renormalization-group framework \cite{Metzner2012} whose main ingredient are coupled flow equations for the single-particle self-energy $\Sigma(\omega)$ as well as for the effective two-body interaction on the Keldysh contour. The fundamental approximation within this method is a truncation of the flow equations in powers of $U$. We keep all terms of $\mathcal{O}(U^2)$ as well as an infinite number of certain higher-order terms; we obtain exact results for $U=0$. The key advantage of our approach is that it keeps track of all microscopic details, which is essential in non-equilibrium where an {\it a priori} separation of relevant low-energy degrees of freedoms from the rest is ambiguous. In contrast to prior, leading-order applications for 1d systems out of equilibrium which essentially mapped the problem an effective \textit{non-interacting} one \cite{Gezzi2007,Jakobs2007,PhysRevB.88.165131}, we incorporate inelastic processes by accounting for the energy-dependence of the effective two-body interaction \cite{Karrasch2008,Jakobs2010b}. The resulting set of RG flow equations is involved, and their solution requires advanced concepts such as recursion relations for Green's functions in the presence of quasi-translation invariance. Details of this method can be found in a separate publication \cite{longpaper}.

In contrast to both mean-field and perturbation theory, our RG approach yields a finite value of $U_\mathrm{crit}$ in the limit $\Gamma=E=0$ in accord with the analytic solution \cite{Markhof2018}. Moreover, we quantitatively reproduce an exact result for the phase boundary at $\Gamma>0$ in the limit $U\to\infty$. Both are highly non-trivial tests for our method and indicate that the phase diagram can be determined reliably.

\begin{figure}[t]
\includegraphics[width=\columnwidth,clip]{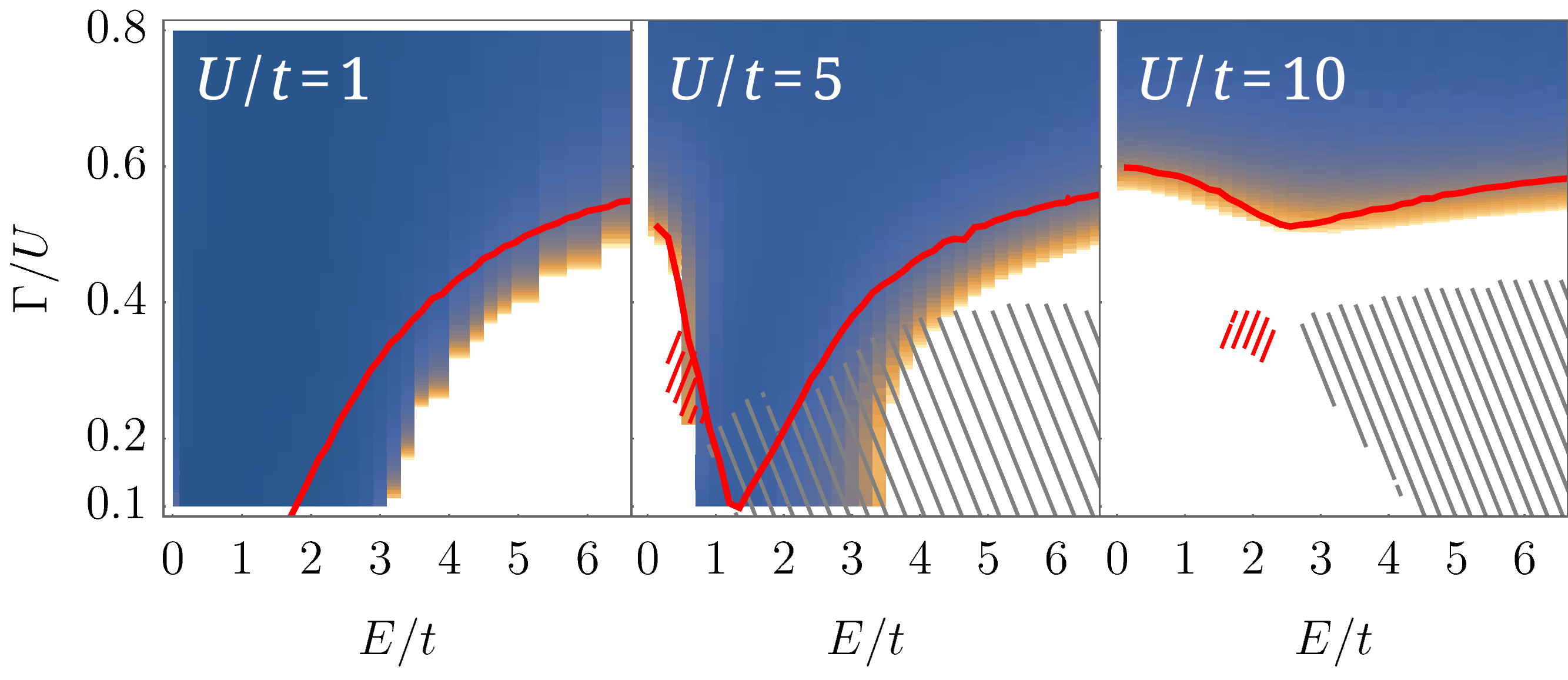}
  \caption{Phase diagram in the $\Gamma-E-$plane for different values of $U$. At intermediate $U/t=5$ and small $\Gamma$, an electric field first induces currents and drives the system into a disordered metallic phase. For large $E$, however, many-body Wannier-Stark localization sets in and the system re-enters a CDW phase. For small $U/t=1$, a large electric field can drive the transition into an ordered CDW phase. At large $U/t=10$, the phase boundary can be extracted reliably via mean-field theory (red line); our RG data is in good agreement with this prediction. Grey hatching indicates regions where a mean-field approach fails to find any stable solution. Red hatching as well as the color scale are the same as in Fig.~\ref{fig:inter_field}}\label{fig:large_field}
  \end{figure}

{\it Phase Diagram--- } To probe spontaneous symmetry breaking and the transition into a CDW phase, we add a small staggered on-site potential $s$ to the Hamiltonian \cite{Markhof2018} and compute the susceptibility
\begin{equation}
\label{eq:chi}
    \chi=\lim_{s\to 0} \frac{\langle\hat n_\mathrm{even}\rangle-\langle\hat n_\mathrm{odd}\rangle}{s},
    \end{equation}
where $\langle \hat n_\text{even}\rangle$ and $\langle \hat n_\text{odd}\rangle$ denote the occupation of even and odd sites, respectively. For systems transitioning into a charge density wave state, $\chi$ will diverge, while it remains finite otherwise. We can thus employ $\chi$ to map out the phase diagram. 

In the left and right panel of Fig.~\ref{fig:inter_field}, we show phase diagrams in the $\Gamma-U-$plane for zero and finite electric fields, respectively. The false color plot shows the susceptibility \eqref{eq:chi} obtained within our second-order RG scheme. 
The red line indicates the mean field transition into a charge-ordered state. In equilibrium ($E=0$) and at low $\Gamma$, we find $U^{\rm RG}_\mathrm{crit}/t\approx 1.4$ in accord with the exact solution and in contrast to both the mean-field ($U^{\rm MF}_\mathrm{crit}/t=0$) and perturbation-theory ($U^{\rm PT}_\mathrm{crit}/t=\infty$) results. If the reservoir coupling $\Gamma$ is increased, the system is ultimately driven into a translation-invariant, disordered phase. The size of the CDW phase decreases if a small electric field is turned on: Currents start to flow and tend to destroy charge order (see right panel of Fig.~\ref{fig:inter_field}).

The mean-field equations can generally have multiple solutions, and in equilibrium the physical one can be determined by minimizing the free energy. At finite electric fields, however, it is unclear how to proceed, and it is not even guaranteed that a single stable mean-field solution exits \cite{PhysRevB.100.121110}. This is explicitly illustrated in Fig.~\ref{fig:inter_field} where the red hatched regions show regimes where at least one solution indicating charge order and at least one stable solution indicating a uniform state can be found. Thus, it is difficult to decide the physically relevant non-equilibrium steady-state using mean-field theory. Such a deficiency does not occur within our RG approach.

In Fig.~\ref{fig:large_field}, we investigate the influence of the electric field (which drives the system into a non-equilibrium state) in more detail and report an astonishingly delicate dependence of the phase boundary on the field strength. The three different panels show the phase diagram in the $\Gamma-E-$plane for three different values of the interaction $U$.  At intermediate interaction $U/t=5$, the system is in a CDW phase for small $\Gamma$  and $E$. If the electric field is increased, currents start to flow and suppress charge order. At large $E$, however, large site-to-site energy gradients energetically prohibit the movement of particles. In this many-body Wannier-Stark localized regime, currents vanish and the system re-enters a CDW phase. For small $U/t=1$, the electric field \textit{drives} the transition into a CDW phase. This illustrates the intricate nature of equilibrium BKT transitions and how their behavior can be flexibly tuned when driving the system into an out-of-equilibrium state. 

If either the interaction or the electric field becomes large, we find good agreement between the mean-field prediction and our RG approach (see Fig.~\ref{fig:large_field}), which is reassuring as quantum fluctuations should be small in these limits. In particular, the mean-field equations can be solved analytically for $U\to\infty$ or $E\to\infty$, and a transition between a charge ordered and a metallic state occurs at a critical value of
$\Gamma^\mathrm{MF}_\mathrm{crit}=\frac{2}{\pi}U^\mathrm{MF}_\mathrm{crit}\approx 0.64U^\mathrm{MF}_\mathrm{crit}
$ \cite{longpaper}.
Our RG scheme yields $\Gamma^\mathrm{RG}_\mathrm{crit}\approx 0.61U^\mathrm{RG}_\mathrm{crit}$. We emphasize that this represents a highly non-trivial test for our framework which is based on a truncation of a hierarchy of flow equations in powers of $U$. One should note that at large but finite $U/t=5$ and $U/t=10$, the RG approach has again a distinct advantage over the mean-field analysis: Within the latter, there are regions in the phase diagram where no stable solution can be obtained (grey hatching in Fig.~\ref{fig:large_field}).

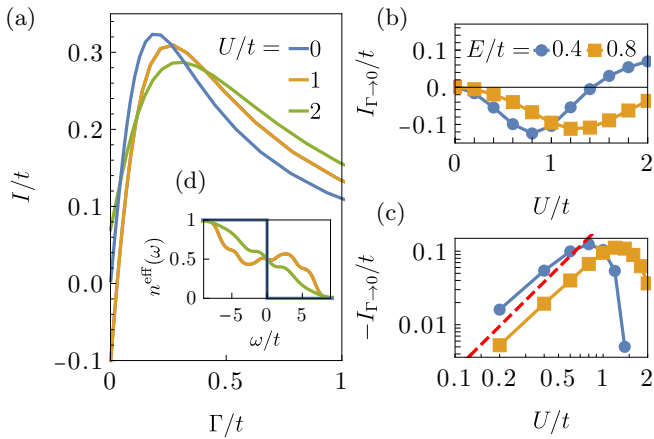
\begin{figure}
    \begin{overpic}[width=\columnwidth]{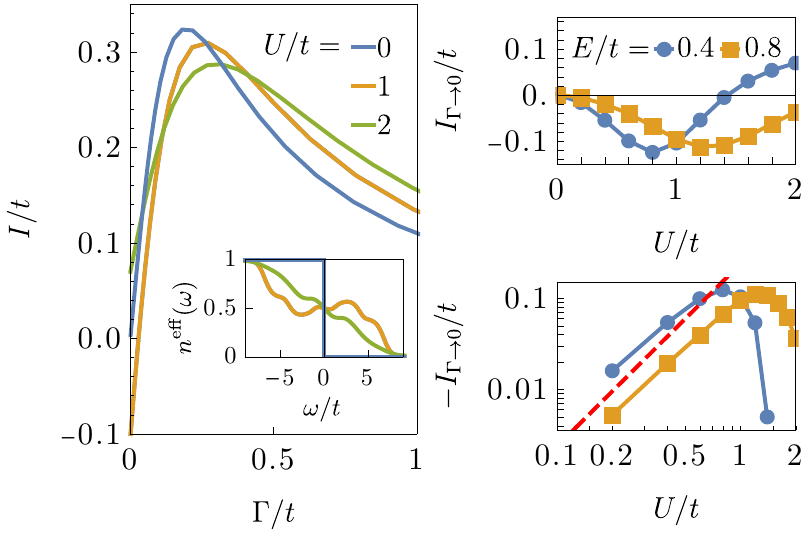}
        \put(0,65){(a)}
        \put(26,40){(d)}
        \put(57,65){(b)}
        \put(57,35){(c)}
    \end{overpic}
    \caption{
        The particle current induced by the electric field in the metallic phase (a) as a function of the reservoir coupling $\Gamma$ for $E/t=0.4$ and different \(U\), 
        (b,c) as a function of the interaction $U$ in the limit $\Gamma\to0$ for different $E$ on a linear and logarithmic scale, respectively. At finite interactions, the current does not vanish in the limit $\Gamma\to0$ and becomes {\it negative} for sufficiently small $U$.
        (d) The effective distribution function (\ref{eq:effdist}) for $E/t=0.4$, $\Gamma\to0$, and various $U$. At $U/t=1$, one observes a positive slope for frequencies smaller than the bandwidth. This corresponds to an {\it effective negative temperature}, which in turn leads to a (negative) current that flows opposite to the direction of the electric field.
    }\label{fig:cur}
\end{figure}

{\it Negative Currents and Temperatures--- }
Next, we study the metallic phase (where no charge order emerges) in more detail. We focus on the limit $\Gamma\to0$, which is particularly interesting as correlations are expected to have the strongest effect on the non-equilibrium steady-state and might lead to exotic phenomena. Fig.~\ref{fig:cur}(a) shows the current flowing through the system as a function of the reservoir coupling $\Gamma$ for three values of the interaction. In the non-interacting case (blue line), the current $I$ first increases as $\Gamma$ is lowered -- particles move with increasing ease without being hampered by the side-coupled reservoirs. At about $\Gamma/t=0.2$, the current features a maximum and decreases as $\Gamma$ is lowered further due to the onset of Wannier-Stark localization. In the presence of interactions, the maximum in the current shifts to larger values of $\Gamma/t$ and the maximal value in $I/t$ decreases. This is most likely related to the interaction-induced band-broadening and to inelastic processes generated by the interaction. The most surprising effect that we report, however, is a {\it negative current} in the steady-state for $\Gamma\to 0$ and small to intermediate $U$ [yellow line in (a)]. Only at larger $U$, the current turns positive again [green line in (a)]. A current which is {\it counter-propagating} to the direction of the electric field \cite{PhysRevLett.91.056803,PhysRevB.71.115316} is striking and in stark contrast to a more commonly encountered negative differential conductance \cite{Taguchi2000,Boulat2008,Inada2009,Mori2009}. In thermal equilibrium, such a state would violate the thermodynamic stability of the system, and spontaneous charge separation would establish a new stable equilibrium configuration. However, no such constrains are present in our generic non-equilibrium setup where energy is not conserved.

It is important to demonstrate that the negative current is not an artifact -- be it numerical or conceptual -- of the underlying RG approach. To this end, we first check that our results are converged w.r.t.~all of the numerical parameters [see the SM for details]. On the conceptual level, we reiterate that our method includes all terms to order $\mathcal{O}(U^2)$ and that possibly uncontrolled corrections can only occur in $\mathcal{O}(U^3)$. Fig.~\ref{fig:cur}(b) and (c) show the current in the limit $\Gamma\to 0$ for different values of  $U/t$. One observes a clear $U^2$-dependence (dashed red line), which is well under control in our approach. This rules out the possibility that the negative current is an artifact of our method.

One can gain some understanding of the counter-intuitive finding of a negative current by considering the local effective distribution function:
\begin{equation}\label{eq:effdist}
    n^{\rm eff}_i(\omega)= \frac12-\frac{{\rm Im}[\Sigma^\mathrm{K}_{ii}(\omega)]}{4{\rm Im}[\Sigma^\mathrm{ret}_{ii}(\omega)]}.
\end{equation}
Due to the combined translation and energy-shift symmetry $n^{\rm eff}_i(\omega)=n^{\rm eff}_{i+j}(\omega+jE)$, we can restrict ourselves to $n^{\rm eff}(\omega)\equiv n^{\rm eff}_0(\omega)$. In equilibrium, our method recovers the dissipation-fluctuation theorem, and Eq.~(\ref{eq:effdist}) reduces to the Fermi-Dirac distribution function. In non-equilibrium, we take $n^{\rm eff}(\omega)$ as a proxy which subsums interaction and driving effects into a distribution function that is effectively imprinted on the different lattice sites. The results are summarized in Fig.~\ref{fig:cur}(d) (we note that this effect cannot be captured by a mean-field treatment). For $U=0$, every site features an effective zero-temperature distribution function imprinted by the reservoirs. However, at finite $U$ this situation changes drastically. At small to moderate $U$ and for frequencies within the bandwidth (those relevant for the transport), the effective distribution function $n^{\rm eff}(\omega)$ inverts its slope (at high frequencies, one recovers the usual Fermi-Dirac behavior). This signals the onset of an {\it effective negative temperature}, and it is well understood that this leads to particles flowing in the opposite, i.e., up-bias direction \cite{Ramsey56,Mosk05,Braun52}.

{\it Conclusion--- } We have studied BKT transitions beyond thermal equilibrium using a novel renormalization-group based framework. Our approach keeps track of all microscopic details during the RG flow, which is necessary in non-equilibrium where an {\it a priori} separation of energy scales is ambiguous. We have applied this machinery to a prototypical model of spinless interacting fermions on a one-dimensional lattice coupled to electronic reservoirs and driven out of equilibrium by an electric field. The closed system is known to exhibit a BKT transition between a metallic and a CDW phase in the equilibrium limit. Our method reproduces this result and also yields quantitative agreement with a mean-field prediction for the phase boundary in limits where quantum fluctuations become irrelevant (e.g., for strong interactions). This shows that our approach is well-suited to determine the non-equilibrium phase diagram for arbitrary parameters. We demonstrated an intricate, highly non-monotonic dependence of the phase boundary on the strength of the electric field, leading to re-enterance behavior. Moreover, we reported on the emergence of a steady-state in (certain parameters regimes of) the open system where a current is flowing up-stream against the drive of the electric field; this state is reminiscent of an equilibrium distribution with an effective negative temperature. Our study can be extended to plethora of questions straightaway such as spinful systems, spin-orbit coupling, topological systems, or systems in two dimensions. Another fruitful direction of future research concerns the question in which sense our results can be interpreted by first mapping the microscopic theory to a continuum field theory Sine-Gordon model, which is known to work in equilibrium. Our results can provide a microscopic testbed to determine under which conditions such a mapping is valid even in the {\it continuously} driven strongly non-equilibrium realm, i.e., beyond the cases of quantum quenches \cite{MatheyL2010,MatheyL2017}. Our approach is also extendable to the case of finite driving frequency of the electric field, which will allow to directly connect to experiments using laser excitations of solids.

\textit{Acknowledgments.} DMK acknowledges funding by the Deutsche Forschungsgemeinschaft (DFG, German Research Foundation) under Germany's Excellence Strategy - Cluster of Excellence Matter and Light for Quantum Computing (ML4Q) EXC 2004/1 - 390534769. We acknowledge support from the Max Planck-New York City Center for Non-Equilibrium Quantum Phenomena. CKa and CKl acknowledge support by the Deutsche Forschungsgemeinschaft through the Emmy Noether program (KA 3360/2-1). CKa acknowledges support by the QuanoMet initiative of Lower Saxony.

\bibliography{ref}

\begin{thebibliography}{40}%
\makeatletter
\providecommand \@ifxundefined [1]{%
 \@ifx{#1\undefined}
}%
\providecommand \@ifnum [1]{%
 \ifnum #1\expandafter \@firstoftwo
 \else \expandafter \@secondoftwo
 \fi
}%
\providecommand \@ifx [1]{%
 \ifx #1\expandafter \@firstoftwo
 \else \expandafter \@secondoftwo
 \fi
}%
\providecommand \natexlab [1]{#1}%
\providecommand \enquote  [1]{``#1''}%
\providecommand \bibnamefont  [1]{#1}%
\providecommand \bibfnamefont [1]{#1}%
\providecommand \citenamefont [1]{#1}%
\providecommand \href@noop [0]{\@secondoftwo}%
\providecommand \href [0]{\begingroup \@sanitize@url \@href}%
\providecommand \@href[1]{\@@startlink{#1}\@@href}%
\providecommand \@@href[1]{\endgroup#1\@@endlink}%
\providecommand \@sanitize@url [0]{\catcode `\\12\catcode `\$12\catcode
  `\&12\catcode `\#12\catcode `\^12\catcode `\_12\catcode `\%12\relax}%
\providecommand \@@startlink[1]{}%
\providecommand \@@endlink[0]{}%
\providecommand \url  [0]{\begingroup\@sanitize@url \@url }%
\providecommand \@url [1]{\endgroup\@href {#1}{\urlprefix }}%
\providecommand \urlprefix  [0]{URL }%
\providecommand \Eprint [0]{\href }%
\providecommand \doibase [0]{http://dx.doi.org/}%
\providecommand \selectlanguage [0]{\@gobble}%
\providecommand \bibinfo  [0]{\@secondoftwo}%
\providecommand \bibfield  [0]{\@secondoftwo}%
\providecommand \translation [1]{[#1]}%
\providecommand \BibitemOpen [0]{}%
\providecommand \bibitemStop [0]{}%
\providecommand \bibitemNoStop [0]{.\EOS\space}%
\providecommand \EOS [0]{\spacefactor3000\relax}%
\providecommand \BibitemShut  [1]{\csname bibitem#1\endcsname}%
\let\auto@bib@innerbib\@empty
\bibitem [{\citenamefont {Basov}\ \emph {et~al.}(2017)\citenamefont {Basov},
  \citenamefont {Averitt},\ and\ \citenamefont {Hsieh}}]{Basov2017}%
  \BibitemOpen
  \bibfield  {author} {\bibinfo {author} {\bibfnamefont {D.~N.}\ \bibnamefont
  {Basov}}, \bibinfo {author} {\bibfnamefont {R.~D.}\ \bibnamefont {Averitt}},
  \ and\ \bibinfo {author} {\bibfnamefont {D.}~\bibnamefont {Hsieh}},\ }\href
  {\doibase 10.1038/nmat5017} {\bibfield  {journal} {\bibinfo  {journal}
  {Nature Materials}\ }\textbf {\bibinfo {volume} {16}},\ \bibinfo {pages}
  {1077} (\bibinfo {year} {2017})}\BibitemShut {NoStop}%
\bibitem [{\citenamefont {Sachdev}(2009)}]{Sachdev2009}%
  \BibitemOpen
  \bibfield  {author} {\bibinfo {author} {\bibfnamefont {S.}~\bibnamefont
  {Sachdev}},\ }\href {\doibase 10.1017/cbo9780511973765} {\emph {\bibinfo
  {title} {Quantum Phase Transitions}}}\ (\bibinfo  {publisher} {Cambridge
  University Press},\ \bibinfo {year} {2009})\BibitemShut {NoStop}%
\bibitem [{\citenamefont {Budden}\ \emph {et~al.}(2020)\citenamefont {Budden},
  \citenamefont {Gebert}, \citenamefont {Buzzi}, \citenamefont {Jotzu},
  \citenamefont {Wang}, \citenamefont {Matsuyama}, \citenamefont {Meier},
  \citenamefont {Laplace}, \citenamefont {Pontiroli}, \citenamefont {Riccò},
  \citenamefont {Schlawin}, \citenamefont {Jaksch},\ and\ \citenamefont
  {Cavalleri}}]{Budden20}%
  \BibitemOpen
  \bibfield  {author} {\bibinfo {author} {\bibfnamefont {M.}~\bibnamefont
  {Budden}}, \bibinfo {author} {\bibfnamefont {T.}~\bibnamefont {Gebert}},
  \bibinfo {author} {\bibfnamefont {M.}~\bibnamefont {Buzzi}}, \bibinfo
  {author} {\bibfnamefont {G.}~\bibnamefont {Jotzu}}, \bibinfo {author}
  {\bibfnamefont {E.}~\bibnamefont {Wang}}, \bibinfo {author} {\bibfnamefont
  {T.}~\bibnamefont {Matsuyama}}, \bibinfo {author} {\bibfnamefont
  {G.}~\bibnamefont {Meier}}, \bibinfo {author} {\bibfnamefont
  {Y.}~\bibnamefont {Laplace}}, \bibinfo {author} {\bibfnamefont
  {D.}~\bibnamefont {Pontiroli}}, \bibinfo {author} {\bibfnamefont
  {M.}~\bibnamefont {Riccò}}, \bibinfo {author} {\bibfnamefont
  {F.}~\bibnamefont {Schlawin}}, \bibinfo {author} {\bibfnamefont
  {D.}~\bibnamefont {Jaksch}}, \ and\ \bibinfo {author} {\bibfnamefont
  {A.}~\bibnamefont {Cavalleri}},\ }\href@noop {} {\enquote {\bibinfo {title}
  {Evidence for metastable photo-induced superconductivity in k3c60},}\ }
  (\bibinfo {year} {2020}),\ \Eprint {http://arxiv.org/abs/arXiv:2002.12835}
  {arXiv:2002.12835} \BibitemShut {NoStop}%
\bibitem [{\citenamefont {Mermin}\ and\ \citenamefont
  {Wagner}(1966)}]{Mermin66}%
  \BibitemOpen
  \bibfield  {author} {\bibinfo {author} {\bibfnamefont {N.~D.}\ \bibnamefont
  {Mermin}}\ and\ \bibinfo {author} {\bibfnamefont {H.}~\bibnamefont
  {Wagner}},\ }\href {\doibase 10.1103/PhysRevLett.17.1133} {\bibfield
  {journal} {\bibinfo  {journal} {Phys. Rev. Lett.}\ }\textbf {\bibinfo
  {volume} {17}},\ \bibinfo {pages} {1133} (\bibinfo {year}
  {1966})}\BibitemShut {NoStop}%
\bibitem [{\citenamefont {Hohenberg}(1967)}]{Hohenberg67}%
  \BibitemOpen
  \bibfield  {author} {\bibinfo {author} {\bibfnamefont {P.~C.}\ \bibnamefont
  {Hohenberg}},\ }\href {\doibase 10.1103/PhysRev.158.383} {\bibfield
  {journal} {\bibinfo  {journal} {Phys. Rev.}\ }\textbf {\bibinfo {volume}
  {158}},\ \bibinfo {pages} {383} (\bibinfo {year} {1967})}\BibitemShut
  {NoStop}%
\bibitem [{\citenamefont {Walldorf}\ \emph {et~al.}(2019)\citenamefont
  {Walldorf}, \citenamefont {Kennes}, \citenamefont {Paaske},\ and\
  \citenamefont {Millis}}]{PhysRevB.100.121110}%
  \BibitemOpen
  \bibfield  {author} {\bibinfo {author} {\bibfnamefont {N.}~\bibnamefont
  {Walldorf}}, \bibinfo {author} {\bibfnamefont {D.~M.}\ \bibnamefont
  {Kennes}}, \bibinfo {author} {\bibfnamefont {J.}~\bibnamefont {Paaske}}, \
  and\ \bibinfo {author} {\bibfnamefont {A.~J.}\ \bibnamefont {Millis}},\
  }\href {\doibase 10.1103/PhysRevB.100.121110} {\bibfield  {journal} {\bibinfo
   {journal} {Phys. Rev. B}\ }\textbf {\bibinfo {volume} {100}},\ \bibinfo
  {pages} {121110} (\bibinfo {year} {2019})}\BibitemShut {NoStop}%
\bibitem [{\citenamefont {Jakobs}\ \emph {et~al.}(2007)\citenamefont {Jakobs},
  \citenamefont {Meden},\ and\ \citenamefont {Schoeller}}]{Jakobs2007}%
  \BibitemOpen
  \bibfield  {author} {\bibinfo {author} {\bibfnamefont {S.~G.}\ \bibnamefont
  {Jakobs}}, \bibinfo {author} {\bibfnamefont {V.}~\bibnamefont {Meden}}, \
  and\ \bibinfo {author} {\bibfnamefont {H.}~\bibnamefont {Schoeller}},\ }\href
  {\doibase 10.1103/PhysRevLett.99.150603} {\bibfield  {journal} {\bibinfo
  {journal} {Phys. Rev. Lett.}\ }\textbf {\bibinfo {volume} {99}},\ \bibinfo
  {pages} {150603} (\bibinfo {year} {2007})}\BibitemShut {NoStop}%
\bibitem [{\citenamefont {Mitra}\ and\ \citenamefont
  {Millis}(2008)}]{Mitra2008}%
  \BibitemOpen
  \bibfield  {author} {\bibinfo {author} {\bibfnamefont {A.}~\bibnamefont
  {Mitra}}\ and\ \bibinfo {author} {\bibfnamefont {A.~J.}\ \bibnamefont
  {Millis}},\ }\href {\doibase 10.1103/PhysRevB.77.220404} {\bibfield
  {journal} {\bibinfo  {journal} {Phys. Rev. B}\ }\textbf {\bibinfo {volume}
  {77}},\ \bibinfo {pages} {220404} (\bibinfo {year} {2008})}\BibitemShut
  {NoStop}%
\bibitem [{\citenamefont {Kennes}\ \emph {et~al.}(2012)\citenamefont {Kennes},
  \citenamefont {Jakobs}, \citenamefont {Karrasch},\ and\ \citenamefont
  {Meden}}]{Kennes2012}%
  \BibitemOpen
  \bibfield  {author} {\bibinfo {author} {\bibfnamefont {D.~M.}\ \bibnamefont
  {Kennes}}, \bibinfo {author} {\bibfnamefont {S.~G.}\ \bibnamefont {Jakobs}},
  \bibinfo {author} {\bibfnamefont {C.}~\bibnamefont {Karrasch}}, \ and\
  \bibinfo {author} {\bibfnamefont {V.}~\bibnamefont {Meden}},\ }\href
  {\doibase 10.1103/PhysRevB.85.085113} {\bibfield  {journal} {\bibinfo
  {journal} {Phys. Rev. B}\ }\textbf {\bibinfo {volume} {85}},\ \bibinfo
  {pages} {085113} (\bibinfo {year} {2012})}\BibitemShut {NoStop}%
\bibitem [{\citenamefont {Aoki}\ \emph {et~al.}(2014)\citenamefont {Aoki},
  \citenamefont {Tsuji}, \citenamefont {Eckstein}, \citenamefont {Kollar},
  \citenamefont {Oka},\ and\ \citenamefont {Werner}}]{Aoki2014}%
  \BibitemOpen
  \bibfield  {author} {\bibinfo {author} {\bibfnamefont {H.}~\bibnamefont
  {Aoki}}, \bibinfo {author} {\bibfnamefont {N.}~\bibnamefont {Tsuji}},
  \bibinfo {author} {\bibfnamefont {M.}~\bibnamefont {Eckstein}}, \bibinfo
  {author} {\bibfnamefont {M.}~\bibnamefont {Kollar}}, \bibinfo {author}
  {\bibfnamefont {T.}~\bibnamefont {Oka}}, \ and\ \bibinfo {author}
  {\bibfnamefont {P.}~\bibnamefont {Werner}},\ }\href {\doibase
  10.1103/revmodphys.86.779} {\bibfield  {journal} {\bibinfo  {journal}
  {Reviews of Modern Physics}\ }\textbf {\bibinfo {volume} {86}},\ \bibinfo
  {pages} {779} (\bibinfo {year} {2014})}\BibitemShut {NoStop}%
\bibitem [{\citenamefont {Sieberer}\ \emph {et~al.}(2016)\citenamefont
  {Sieberer}, \citenamefont {Buchhold},\ and\ \citenamefont
  {Diehl}}]{Sieberer2016}%
  \BibitemOpen
  \bibfield  {author} {\bibinfo {author} {\bibfnamefont {L.~M.}\ \bibnamefont
  {Sieberer}}, \bibinfo {author} {\bibfnamefont {M.}~\bibnamefont {Buchhold}},
  \ and\ \bibinfo {author} {\bibfnamefont {S.}~\bibnamefont {Diehl}},\ }\href
  {\doibase 10.1088/0034-4885/79/9/096001} {\bibfield  {journal} {\bibinfo
  {journal} {Reports on Progress in Physics}\ }\textbf {\bibinfo {volume}
  {79}},\ \bibinfo {pages} {096001} (\bibinfo {year} {2016})}\BibitemShut
  {NoStop}%
\bibitem [{\citenamefont {Eissing}\ \emph {et~al.}(2016)\citenamefont
  {Eissing}, \citenamefont {Meden},\ and\ \citenamefont {Kennes}}]{AK1}%
  \BibitemOpen
  \bibfield  {author} {\bibinfo {author} {\bibfnamefont {A.~K.}\ \bibnamefont
  {Eissing}}, \bibinfo {author} {\bibfnamefont {V.}~\bibnamefont {Meden}}, \
  and\ \bibinfo {author} {\bibfnamefont {D.~M.}\ \bibnamefont {Kennes}},\
  }\href {\doibase 10.1103/PhysRevLett.116.026801} {\bibfield  {journal}
  {\bibinfo  {journal} {Phys. Rev. Lett.}\ }\textbf {\bibinfo {volume} {116}},\
  \bibinfo {pages} {026801} (\bibinfo {year} {2016})}\BibitemShut {NoStop}%
\bibitem [{\citenamefont {Kennes}\ \emph {et~al.}(2018)\citenamefont {Kennes},
  \citenamefont {de~la Torre}, \citenamefont {Ron}, \citenamefont {Hsieh},\
  and\ \citenamefont {Millis}}]{PhysRevLett.120.127601}%
  \BibitemOpen
  \bibfield  {author} {\bibinfo {author} {\bibfnamefont {D.~M.}\ \bibnamefont
  {Kennes}}, \bibinfo {author} {\bibfnamefont {A.}~\bibnamefont {de~la Torre}},
  \bibinfo {author} {\bibfnamefont {A.}~\bibnamefont {Ron}}, \bibinfo {author}
  {\bibfnamefont {D.}~\bibnamefont {Hsieh}}, \ and\ \bibinfo {author}
  {\bibfnamefont {A.~J.}\ \bibnamefont {Millis}},\ }\href {\doibase
  10.1103/PhysRevLett.120.127601} {\bibfield  {journal} {\bibinfo  {journal}
  {Phys. Rev. Lett.}\ }\textbf {\bibinfo {volume} {120}},\ \bibinfo {pages}
  {127601} (\bibinfo {year} {2018})}\BibitemShut {NoStop}%
\bibitem [{\citenamefont {Fisher}(1974)}]{RevModPhys.46.597}%
  \BibitemOpen
  \bibfield  {author} {\bibinfo {author} {\bibfnamefont {M.~E.}\ \bibnamefont
  {Fisher}},\ }\href {\doibase 10.1103/RevModPhys.46.597} {\bibfield  {journal}
  {\bibinfo  {journal} {Rev. Mod. Phys.}\ }\textbf {\bibinfo {volume} {46}},\
  \bibinfo {pages} {597} (\bibinfo {year} {1974})}\BibitemShut {NoStop}%
\bibitem [{\citenamefont {Wilson}(1975)}]{RevModPhys.47.773}%
  \BibitemOpen
  \bibfield  {author} {\bibinfo {author} {\bibfnamefont {K.~G.}\ \bibnamefont
  {Wilson}},\ }\href {\doibase 10.1103/RevModPhys.47.773} {\bibfield  {journal}
  {\bibinfo  {journal} {Rev. Mod. Phys.}\ }\textbf {\bibinfo {volume} {47}},\
  \bibinfo {pages} {773} (\bibinfo {year} {1975})}\BibitemShut {NoStop}%
\bibitem [{\citenamefont {Metzner}\ \emph {et~al.}(2012)\citenamefont
  {Metzner}, \citenamefont {Salmhofer}, \citenamefont {Honerkamp},
  \citenamefont {Meden},\ and\ \citenamefont {Sch\"{o}nhammer}}]{Metzner2012}%
  \BibitemOpen
  \bibfield  {author} {\bibinfo {author} {\bibfnamefont {W.}~\bibnamefont
  {Metzner}}, \bibinfo {author} {\bibfnamefont {M.}~\bibnamefont {Salmhofer}},
  \bibinfo {author} {\bibfnamefont {C.}~\bibnamefont {Honerkamp}}, \bibinfo
  {author} {\bibfnamefont {V.}~\bibnamefont {Meden}}, \ and\ \bibinfo {author}
  {\bibfnamefont {K.}~\bibnamefont {Sch\"{o}nhammer}},\ }\href {\doibase
  10.1103/revmodphys.84.299} {\bibfield  {journal} {\bibinfo  {journal}
  {Reviews of Modern Physics}\ }\textbf {\bibinfo {volume} {84}},\ \bibinfo
  {pages} {299} (\bibinfo {year} {2012})}\BibitemShut {NoStop}%
\bibitem [{\citenamefont {Kopietz}\ \emph {et~al.}(2010)\citenamefont
  {Kopietz}, \citenamefont {Bartosch},\ and\ \citenamefont
  {Sch\"utz}}]{kopietzBook}%
  \BibitemOpen
  \bibfield  {author} {\bibinfo {author} {\bibfnamefont {P.}~\bibnamefont
  {Kopietz}}, \bibinfo {author} {\bibfnamefont {L.}~\bibnamefont {Bartosch}}, \
  and\ \bibinfo {author} {\bibfnamefont {F.}~\bibnamefont {Sch\"utz}},\ }\href
  {\doibase 10.1007/978-3-642-05094-7} {\emph {\bibinfo {title} {Introduction
  to the Functional Renormalization Group}}},\ Lecture Notes in Physics\
  (\bibinfo  {publisher} {Springer-Verlag Berlin Heidelberg},\ \bibinfo {year}
  {2010})\BibitemShut {NoStop}%
\bibitem [{\citenamefont {Mathey}\ and\ \citenamefont
  {Polkovnikov}(2010)}]{MatheyL2010}%
  \BibitemOpen
  \bibfield  {author} {\bibinfo {author} {\bibfnamefont {L.}~\bibnamefont
  {Mathey}}\ and\ \bibinfo {author} {\bibfnamefont {A.}~\bibnamefont
  {Polkovnikov}},\ }\href {\doibase 10.1103/PhysRevA.81.033605} {\bibfield
  {journal} {\bibinfo  {journal} {Phys. Rev. A}\ }\textbf {\bibinfo {volume}
  {81}},\ \bibinfo {pages} {033605} (\bibinfo {year} {2010})}\BibitemShut
  {NoStop}%
\bibitem [{\citenamefont {Mathey}\ \emph {et~al.}(2017)\citenamefont {Mathey},
  \citenamefont {G\"unter}, \citenamefont {Dalibard},\ and\ \citenamefont
  {Polkovnikov}}]{MatheyL2017}%
  \BibitemOpen
  \bibfield  {author} {\bibinfo {author} {\bibfnamefont {L.}~\bibnamefont
  {Mathey}}, \bibinfo {author} {\bibfnamefont {K.~J.}\ \bibnamefont
  {G\"unter}}, \bibinfo {author} {\bibfnamefont {J.}~\bibnamefont {Dalibard}},
  \ and\ \bibinfo {author} {\bibfnamefont {A.}~\bibnamefont {Polkovnikov}},\
  }\href {\doibase 10.1103/PhysRevA.95.053630} {\bibfield  {journal} {\bibinfo
  {journal} {Phys. Rev. A}\ }\textbf {\bibinfo {volume} {95}},\ \bibinfo
  {pages} {053630} (\bibinfo {year} {2017})}\BibitemShut {NoStop}%
\bibitem [{\citenamefont {Kosterlitz}\ and\ \citenamefont
  {Thouless}(1973)}]{Kosterlitz_1973}%
  \BibitemOpen
  \bibfield  {author} {\bibinfo {author} {\bibfnamefont {J.~M.}\ \bibnamefont
  {Kosterlitz}}\ and\ \bibinfo {author} {\bibfnamefont {D.~J.}\ \bibnamefont
  {Thouless}},\ }\href {\doibase 10.1088/0022-3719/6/7/010} {\bibfield
  {journal} {\bibinfo  {journal} {Journal of Physics C: Solid State Physics}\
  }\textbf {\bibinfo {volume} {6}},\ \bibinfo {pages} {1181} (\bibinfo {year}
  {1973})}\BibitemShut {NoStop}%
\bibitem [{\citenamefont {Giamarchi}(2004)}]{giamarchi}%
  \BibitemOpen
  \bibfield  {author} {\bibinfo {author} {\bibfnamefont {T.}~\bibnamefont
  {Giamarchi}},\ }\href@noop {} {\emph {\bibinfo {title} {Quantum {P}hysics in
  {O}ne {D}imension}}}\ (\bibinfo  {publisher} {Clarendon Press},\ \bibinfo
  {address} {Oxford},\ \bibinfo {year} {2004})\ p.\ \bibinfo {pages}
  {2905}\BibitemShut {NoStop}%
\bibitem [{\citenamefont {Wannier}(1962)}]{Wannier1962}%
  \BibitemOpen
  \bibfield  {author} {\bibinfo {author} {\bibfnamefont {G.~H.}\ \bibnamefont
  {Wannier}},\ }\href {\doibase 10.1103/revmodphys.34.645} {\bibfield
  {journal} {\bibinfo  {journal} {Reviews of Modern Physics}\ }\textbf
  {\bibinfo {volume} {34}},\ \bibinfo {pages} {645} (\bibinfo {year}
  {1962})}\BibitemShut {NoStop}%
\bibitem [{\citenamefont {van Nieuwenburg}\ \emph {et~al.}(2019)\citenamefont
  {van Nieuwenburg}, \citenamefont {Baum},\ and\ \citenamefont
  {Refael}}]{Stark1}%
  \BibitemOpen
  \bibfield  {author} {\bibinfo {author} {\bibfnamefont {E.}~\bibnamefont {van
  Nieuwenburg}}, \bibinfo {author} {\bibfnamefont {Y.}~\bibnamefont {Baum}}, \
  and\ \bibinfo {author} {\bibfnamefont {G.}~\bibnamefont {Refael}},\ }\href
  {\doibase 10.1073/pnas.1819316116} {\bibfield  {journal} {\bibinfo  {journal}
  {PNAS}\ }\textbf {\bibinfo {volume} {116}},\ \bibinfo {pages} {9269}
  (\bibinfo {year} {2019})}\BibitemShut {NoStop}%
\bibitem [{\citenamefont {Schulz}\ \emph {et~al.}(2019)\citenamefont {Schulz},
  \citenamefont {Hooley}, \citenamefont {Moessner},\ and\ \citenamefont
  {Pollmann}}]{Stark2}%
  \BibitemOpen
  \bibfield  {author} {\bibinfo {author} {\bibfnamefont {M.}~\bibnamefont
  {Schulz}}, \bibinfo {author} {\bibfnamefont {C.~A.}\ \bibnamefont {Hooley}},
  \bibinfo {author} {\bibfnamefont {R.}~\bibnamefont {Moessner}}, \ and\
  \bibinfo {author} {\bibfnamefont {F.}~\bibnamefont {Pollmann}},\ }\href
  {\doibase 10.1103/PhysRevLett.122.040606} {\bibfield  {journal} {\bibinfo
  {journal} {Phys. Rev. Lett.}\ }\textbf {\bibinfo {volume} {122}},\ \bibinfo
  {pages} {040606} (\bibinfo {year} {2019})}\BibitemShut {NoStop}%
\bibitem [{\citenamefont {Ramsey}(1956)}]{Ramsey56}%
  \BibitemOpen
  \bibfield  {author} {\bibinfo {author} {\bibfnamefont {N.~F.}\ \bibnamefont
  {Ramsey}},\ }\href {\doibase 10.1103/PhysRev.103.20} {\bibfield  {journal}
  {\bibinfo  {journal} {Phys. Rev.}\ }\textbf {\bibinfo {volume} {103}},\
  \bibinfo {pages} {20} (\bibinfo {year} {1956})}\BibitemShut {NoStop}%
\bibitem [{\citenamefont {Mosk}(2005)}]{Mosk05}%
  \BibitemOpen
  \bibfield  {author} {\bibinfo {author} {\bibfnamefont {A.~P.}\ \bibnamefont
  {Mosk}},\ }\href {\doibase 10.1103/PhysRevLett.95.040403} {\bibfield
  {journal} {\bibinfo  {journal} {Phys. Rev. Lett.}\ }\textbf {\bibinfo
  {volume} {95}},\ \bibinfo {pages} {040403} (\bibinfo {year}
  {2005})}\BibitemShut {NoStop}%
\bibitem [{\citenamefont {Braun}\ \emph {et~al.}(2013)\citenamefont {Braun},
  \citenamefont {Ronzheimer}, \citenamefont {Schreiber}, \citenamefont
  {Hodgman}, \citenamefont {Rom}, \citenamefont {Bloch},\ and\ \citenamefont
  {Schneider}}]{Braun52}%
  \BibitemOpen
  \bibfield  {author} {\bibinfo {author} {\bibfnamefont {S.}~\bibnamefont
  {Braun}}, \bibinfo {author} {\bibfnamefont {J.~P.}\ \bibnamefont
  {Ronzheimer}}, \bibinfo {author} {\bibfnamefont {M.}~\bibnamefont
  {Schreiber}}, \bibinfo {author} {\bibfnamefont {S.~S.}\ \bibnamefont
  {Hodgman}}, \bibinfo {author} {\bibfnamefont {T.}~\bibnamefont {Rom}},
  \bibinfo {author} {\bibfnamefont {I.}~\bibnamefont {Bloch}}, \ and\ \bibinfo
  {author} {\bibfnamefont {U.}~\bibnamefont {Schneider}},\ }\href {\doibase
  10.1126/science.1227831} {\bibfield  {journal} {\bibinfo  {journal}
  {Science}\ }\textbf {\bibinfo {volume} {339}},\ \bibinfo {pages} {52}
  (\bibinfo {year} {2013})}\BibitemShut {NoStop}%
\bibitem [{\citenamefont {Andreev}\ \emph {et~al.}(2003)\citenamefont
  {Andreev}, \citenamefont {Aleiner},\ and\ \citenamefont
  {Millis}}]{PhysRevLett.91.056803}%
  \BibitemOpen
  \bibfield  {author} {\bibinfo {author} {\bibfnamefont {A.~V.}\ \bibnamefont
  {Andreev}}, \bibinfo {author} {\bibfnamefont {I.~L.}\ \bibnamefont
  {Aleiner}}, \ and\ \bibinfo {author} {\bibfnamefont {A.~J.}\ \bibnamefont
  {Millis}},\ }\href {\doibase 10.1103/PhysRevLett.91.056803} {\bibfield
  {journal} {\bibinfo  {journal} {Phys. Rev. Lett.}\ }\textbf {\bibinfo
  {volume} {91}},\ \bibinfo {pages} {056803} (\bibinfo {year}
  {2003})}\BibitemShut {NoStop}%
\bibitem [{\citenamefont {Dmitriev}\ \emph {et~al.}(2005)\citenamefont
  {Dmitriev}, \citenamefont {Vavilov}, \citenamefont {Aleiner}, \citenamefont
  {Mirlin},\ and\ \citenamefont {Polyakov}}]{PhysRevB.71.115316}%
  \BibitemOpen
  \bibfield  {author} {\bibinfo {author} {\bibfnamefont {I.~A.}\ \bibnamefont
  {Dmitriev}}, \bibinfo {author} {\bibfnamefont {M.~G.}\ \bibnamefont
  {Vavilov}}, \bibinfo {author} {\bibfnamefont {I.~L.}\ \bibnamefont
  {Aleiner}}, \bibinfo {author} {\bibfnamefont {A.~D.}\ \bibnamefont {Mirlin}},
  \ and\ \bibinfo {author} {\bibfnamefont {D.~G.}\ \bibnamefont {Polyakov}},\
  }\href {\doibase 10.1103/PhysRevB.71.115316} {\bibfield  {journal} {\bibinfo
  {journal} {Phys. Rev. B}\ }\textbf {\bibinfo {volume} {71}},\ \bibinfo
  {pages} {115316} (\bibinfo {year} {2005})}\BibitemShut {NoStop}%
\bibitem [{\citenamefont {Klöckner}\ \emph {et~al.}(2020)\citenamefont
  {Klöckner}, \citenamefont {Kennes},\ and\ \citenamefont
  {Karrasch}}]{longpaper}%
  \BibitemOpen
  \bibfield  {author} {\bibinfo {author} {\bibfnamefont {C.}~\bibnamefont
  {Klöckner}}, \bibinfo {author} {\bibfnamefont {D.~M.}\ \bibnamefont
  {Kennes}}, \ and\ \bibinfo {author} {\bibfnamefont {C.}~\bibnamefont
  {Karrasch}},\ }\href {\doibase 10.1088/1367-2630/ab990d} {\bibfield
  {journal} {\bibinfo  {journal} {New Journal of Physics}\ }\textbf {\bibinfo
  {volume} {22}},\ \bibinfo {pages} {083039} (\bibinfo {year}
  {2020})}\BibitemShut {NoStop}%
\bibitem [{\citenamefont {Kamenev}\ and\ \citenamefont
  {Levchenko}(2009)}]{Kamenevbook}%
  \BibitemOpen
  \bibfield  {author} {\bibinfo {author} {\bibfnamefont {A.}~\bibnamefont
  {Kamenev}}\ and\ \bibinfo {author} {\bibfnamefont {A.}~\bibnamefont
  {Levchenko}},\ }\href {\doibase 10.1080/00018730902850504} {\bibfield
  {journal} {\bibinfo  {journal} {Advances in Physics}\ }\textbf {\bibinfo
  {volume} {58}},\ \bibinfo {pages} {197} (\bibinfo {year} {2009})}\BibitemShut
  {NoStop}%
\bibitem [{\citenamefont {Gezzi}\ \emph {et~al.}(2007)\citenamefont {Gezzi},
  \citenamefont {Pruschke},\ and\ \citenamefont {Meden}}]{Gezzi2007}%
  \BibitemOpen
  \bibfield  {author} {\bibinfo {author} {\bibfnamefont {R.}~\bibnamefont
  {Gezzi}}, \bibinfo {author} {\bibfnamefont {T.}~\bibnamefont {Pruschke}}, \
  and\ \bibinfo {author} {\bibfnamefont {V.}~\bibnamefont {Meden}},\ }\href
  {\doibase 10.1103/PhysRevB.75.045324} {\bibfield  {journal} {\bibinfo
  {journal} {Phys. Rev. B}\ }\textbf {\bibinfo {volume} {75}},\ \bibinfo
  {pages} {045324} (\bibinfo {year} {2007})}\BibitemShut {NoStop}%
\bibitem [{\citenamefont {Kennes}\ and\ \citenamefont
  {Meden}(2013)}]{PhysRevB.88.165131}%
  \BibitemOpen
  \bibfield  {author} {\bibinfo {author} {\bibfnamefont {D.~M.}\ \bibnamefont
  {Kennes}}\ and\ \bibinfo {author} {\bibfnamefont {V.}~\bibnamefont {Meden}},\
  }\href {\doibase 10.1103/PhysRevB.88.165131} {\bibfield  {journal} {\bibinfo
  {journal} {Phys. Rev. B}\ }\textbf {\bibinfo {volume} {88}},\ \bibinfo
  {pages} {165131} (\bibinfo {year} {2013})}\BibitemShut {NoStop}%
\bibitem [{\citenamefont {Karrasch}\ \emph {et~al.}(2008)\citenamefont
  {Karrasch}, \citenamefont {Hedden}, \citenamefont {Peters}, \citenamefont
  {Pruschke}, \citenamefont {Schönhammer},\ and\ \citenamefont
  {Meden}}]{Karrasch2008}%
  \BibitemOpen
  \bibfield  {author} {\bibinfo {author} {\bibfnamefont {C.}~\bibnamefont
  {Karrasch}}, \bibinfo {author} {\bibfnamefont {R.}~\bibnamefont {Hedden}},
  \bibinfo {author} {\bibfnamefont {R.}~\bibnamefont {Peters}}, \bibinfo
  {author} {\bibfnamefont {T.}~\bibnamefont {Pruschke}}, \bibinfo {author}
  {\bibfnamefont {K.}~\bibnamefont {Schönhammer}}, \ and\ \bibinfo {author}
  {\bibfnamefont {V.}~\bibnamefont {Meden}},\ }\href {\doibase
  10.1088/0953-8984/20/34/345205} {\bibfield  {journal} {\bibinfo  {journal}
  {Journal of Physics: Condensed Matter}\ }\textbf {\bibinfo {volume} {20}},\
  \bibinfo {pages} {345205} (\bibinfo {year} {2008})}\BibitemShut {NoStop}%
\bibitem [{\citenamefont {Jakobs}\ \emph {et~al.}(2010)\citenamefont {Jakobs},
  \citenamefont {Pletyukhov},\ and\ \citenamefont {Schoeller}}]{Jakobs2010b}%
  \BibitemOpen
  \bibfield  {author} {\bibinfo {author} {\bibfnamefont {S.~G.}\ \bibnamefont
  {Jakobs}}, \bibinfo {author} {\bibfnamefont {M.}~\bibnamefont {Pletyukhov}},
  \ and\ \bibinfo {author} {\bibfnamefont {H.}~\bibnamefont {Schoeller}},\
  }\href {\doibase 10.1103/PhysRevB.81.195109} {\bibfield  {journal} {\bibinfo
  {journal} {Phys. Rev. B}\ }\textbf {\bibinfo {volume} {81}},\ \bibinfo
  {pages} {195109} (\bibinfo {year} {2010})}\BibitemShut {NoStop}%
\bibitem [{\citenamefont {Markhof}\ \emph {et~al.}(2018)\citenamefont
  {Markhof}, \citenamefont {Sbierski}, \citenamefont {Meden},\ and\
  \citenamefont {Karrasch}}]{Markhof2018}%
  \BibitemOpen
  \bibfield  {author} {\bibinfo {author} {\bibfnamefont {L.}~\bibnamefont
  {Markhof}}, \bibinfo {author} {\bibfnamefont {B.}~\bibnamefont {Sbierski}},
  \bibinfo {author} {\bibfnamefont {V.}~\bibnamefont {Meden}}, \ and\ \bibinfo
  {author} {\bibfnamefont {C.}~\bibnamefont {Karrasch}},\ }\href {\doibase
  10.1103/PhysRevB.97.235126} {\bibfield  {journal} {\bibinfo  {journal} {Phys.
  Rev. B}\ }\textbf {\bibinfo {volume} {97}},\ \bibinfo {pages} {235126}
  (\bibinfo {year} {2018})}\BibitemShut {NoStop}%
\bibitem [{\citenamefont {Taguchi}\ \emph {et~al.}(2000)\citenamefont
  {Taguchi}, \citenamefont {Matsumoto},\ and\ \citenamefont
  {Tokura}}]{Taguchi2000}%
  \BibitemOpen
  \bibfield  {author} {\bibinfo {author} {\bibfnamefont {Y.}~\bibnamefont
  {Taguchi}}, \bibinfo {author} {\bibfnamefont {T.}~\bibnamefont {Matsumoto}},
  \ and\ \bibinfo {author} {\bibfnamefont {Y.}~\bibnamefont {Tokura}},\ }\href
  {\doibase 10.1103/physrevb.62.7015} {\bibfield  {journal} {\bibinfo
  {journal} {Physical Review B}\ }\textbf {\bibinfo {volume} {62}},\ \bibinfo
  {pages} {7015} (\bibinfo {year} {2000})}\BibitemShut {NoStop}%
\bibitem [{\citenamefont {Boulat}\ \emph {et~al.}(2008)\citenamefont {Boulat},
  \citenamefont {Saleur},\ and\ \citenamefont {Schmitteckert}}]{Boulat2008}%
  \BibitemOpen
  \bibfield  {author} {\bibinfo {author} {\bibfnamefont {E.}~\bibnamefont
  {Boulat}}, \bibinfo {author} {\bibfnamefont {H.}~\bibnamefont {Saleur}}, \
  and\ \bibinfo {author} {\bibfnamefont {P.}~\bibnamefont {Schmitteckert}},\
  }\href {\doibase 10.1103/PhysRevLett.101.140601} {\bibfield  {journal}
  {\bibinfo  {journal} {Phys. Rev. Lett.}\ }\textbf {\bibinfo {volume} {101}},\
  \bibinfo {pages} {140601} (\bibinfo {year} {2008})}\BibitemShut {NoStop}%
\bibitem [{\citenamefont {Inada}\ \emph {et~al.}(2009)\citenamefont {Inada},
  \citenamefont {Terasaki}, \citenamefont {Mori},\ and\ \citenamefont
  {Mori}}]{Inada2009}%
  \BibitemOpen
  \bibfield  {author} {\bibinfo {author} {\bibfnamefont {T.~S.}\ \bibnamefont
  {Inada}}, \bibinfo {author} {\bibfnamefont {I.}~\bibnamefont {Terasaki}},
  \bibinfo {author} {\bibfnamefont {H.}~\bibnamefont {Mori}}, \ and\ \bibinfo
  {author} {\bibfnamefont {T.}~\bibnamefont {Mori}},\ }\href {\doibase
  10.1103/PhysRevB.79.165102} {\bibfield  {journal} {\bibinfo  {journal} {Phys.
  Rev. B}\ }\textbf {\bibinfo {volume} {79}},\ \bibinfo {pages} {165102}
  (\bibinfo {year} {2009})}\BibitemShut {NoStop}%
\bibitem [{\citenamefont {Mori}\ \emph {et~al.}(2009)\citenamefont {Mori},
  \citenamefont {Ozawa}, \citenamefont {Bando}, \citenamefont {Kawamoto},
  \citenamefont {Niizeki}, \citenamefont {Mori},\ and\ \citenamefont
  {Terasaki}}]{Mori2009}%
  \BibitemOpen
  \bibfield  {author} {\bibinfo {author} {\bibfnamefont {T.}~\bibnamefont
  {Mori}}, \bibinfo {author} {\bibfnamefont {T.}~\bibnamefont {Ozawa}},
  \bibinfo {author} {\bibfnamefont {Y.}~\bibnamefont {Bando}}, \bibinfo
  {author} {\bibfnamefont {T.}~\bibnamefont {Kawamoto}}, \bibinfo {author}
  {\bibfnamefont {S.}~\bibnamefont {Niizeki}}, \bibinfo {author} {\bibfnamefont
  {H.}~\bibnamefont {Mori}}, \ and\ \bibinfo {author} {\bibfnamefont
  {I.}~\bibnamefont {Terasaki}},\ }\href {\doibase 10.1103/PhysRevB.79.115108}
  {\bibfield  {journal} {\bibinfo  {journal} {Phys. Rev. B}\ }\textbf {\bibinfo
  {volume} {79}},\ \bibinfo {pages} {115108} (\bibinfo {year}
  {2009})}\BibitemShut {NoStop}%
\end{thebibliography}%

\cleardoublepage

\begin{center}
\bfseries SUPPLEMENTARY INFORMATION
\end{center}

\section{Numerical convergence }

\begin{figure}[h]
    \begin{overpic}[width=\columnwidth]{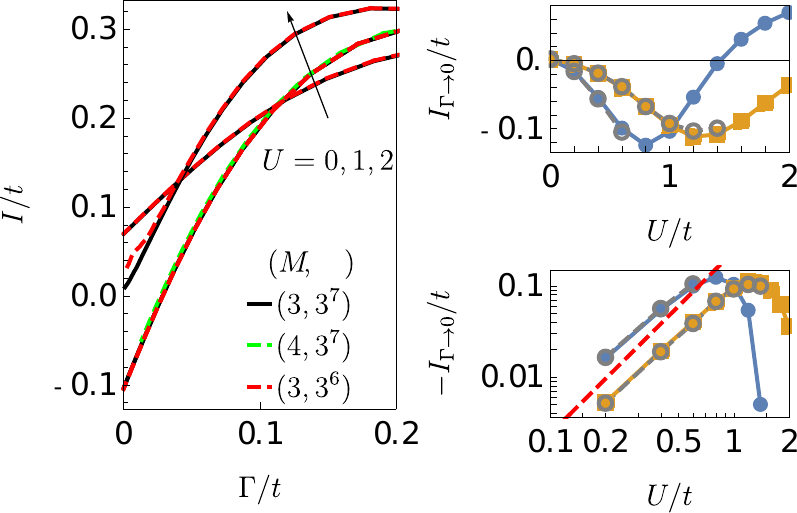}
        \put(0,65){(a)}
        \put(57,65){(b)}
        \put(57,35){(c)}
        \put(39.5,30.25){$N$}
    \end{overpic}
    \caption{
        Demonstration of numerical convergence of the results presented in Fig.~4 of the main text.
         (a) Shows the current varying the correlation length \(M\) and the frequency discretization set by $N$. Only in the absence of scattering (\(U=0\)) we observe some dependence on the chosen frequency discretization, which is the numerically most challenging, but physically trivial parameter set. In addition to the results shown also in Fig.~4 of the main text, here (b) and (c) show the results obtained using \(M=4\) in gray. The negative \(\ord{U^2}\) correction to the current is unaffected.
    }\label{fig:conv}
\end{figure}
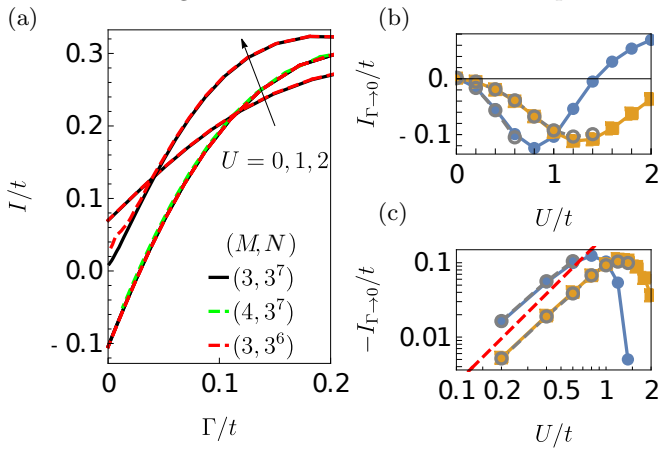

Within the numerical implementation of the renormalization-group framework we present, numerical convergence was always checked. The major parameters, that have to be controlled are (a) the discretization of frequency space and (b) the maximum correlation length \(M\) allowed for the self-energy, enforced by 
\begin{equation}
    \Sigma_{ij}(\omega)\approx 0\ \forall |i-j|\geq M.
\end{equation}
To check convergence with respect to both we vary the size of the frequency grid, where the total number of grid points is denoted by $N$ as well as the length $M$.
 Fig.~\ref{fig:conv} summarizes such convergence checks for the results discussed in Fig.~4 of the main text as an example.

\end{document}